\documentclass[conference]{IEEEtran}
\usepackage{cite}
\usepackage{amsmath,amssymb,amsfonts}
\usepackage{algorithmic}
\usepackage{subcaption}
\usepackage{graphicx}
\usepackage{textcomp}
\usepackage{xcolor}
\usepackage{wasysym}
\usepackage{listings}
\usepackage{float}
\usepackage{url}

\def\BibTeX{{\rm B\kern-.05em{\sc i\kern-.025em b}\kern-.08em
    T\kern-.1667em\lower.7ex\hbox{E}\kern-.125emX}}

\begin{document}
\newcommand{\smartpara}[1]{\noindent \textbf{#1.}}

\newcommand{\RA}[1]{{\color[HTML]{0088AA}{#1}}}
\newcommand{\RB}[1]{{\color[HTML]{AA4499}{#1}}}
\newcommand{\RC}[1]{{\color[HTML]{D55E00}{#1}}}
\newcommand{\RAll}[1]{{\color[HTML]{117733}{#1}}}

\colorlet{myblue}{blue!50!black}
\colorlet{mygreen}{green!50!black}

\newcommand{\xir}{TondIR}
\newcommand{\system}{PyTond}
\newcommand{\decorator}{\code{@pytond}} 

\newcommand{\amirsh}[1]{{\color{red}Amir: #1}}

\newcommand{\hesam}[1]{{\color{red}Hesam: #1}}

\newcommand{\amirali}[1]{{\color{red}Amirali: #1}}

\newcommand{\changed}[1]{{\color{blue}#1}}

\newcommand{\codekwstyle}{\footnotesize\ttfamily\bfseries\color{myblue}}

\newcommand{\code}[1]{{\footnotesize\texttt{#1}}}
\newcommand{\codekw}[1]{{\codekwstyle\texttt{#1}}}
\newcommand{\codetiny}[1]{{\footnotesize\ttfamily\texttt{#1}}}
\newcommand{\codekwpy}[1]{{\footnotesize\color{blue}#1}}

\definecolor{colg}{rgb}{0.1,0.7,0.1}
\definecolor{colr}{rgb}{0.7,0.1,0.1}
\definecolor{colb}{rgb}{0.1,0.1,0.7}
\definecolor{colbb}{rgb}{0.1,0.1,0.6}
\newcommand{\supfull}{{\color{colg}{\CIRCLE}}}
\newcommand{\suphalf}{{\color{colg}{\LEFTcircle}}}
\newcommand{\supnone}{\Circle}

\lstset{emph={head, unique, sort_values, apply, aggregate, groupby, agg, merge, isin, pivot_table, DataFrame, einsum, import, array, all, nonzero, round, max, compress, sum, inner, outer, transpose, matmul}, emphstyle={\color{blue}}, breaklines=true,  basicstyle=\footnotesize}

\lstdefinelanguage{tondir}{
  morekeywords={group, sort, limit, exists,if, sum, UID, True, False, round, min, max},%
  sensitive,%
  morecomment=[l]//,%
  morecomment=[s]{\{}{\}},%
  morestring=[b]",%
  % morestring=[b]',%
  showstringspaces=false,%
  breaklines=true,%
  mathescape=true,%
  showspaces=false,
  showtabs=false,
  showstringspaces=false,
  breakatwhitespace=true,
  xleftmargin=1em,
  % columns=[c]fixed,%
  % basewidth={0.5em, 0.40em},%
  aboveskip=1pt,%\smallskipamount,
  belowskip=1pt,%\negsmallskipamount,
  lineskip=-0.2pt,
%  backgroundcolor=\color{listingbg},
%  basicstyle=\linespread{0.4}\footnotesize\ttfamily,
   % numbers=left,
   % numbersep=5pt,
   % numberstyle=\tiny\ttfamily,
  basicstyle=\footnotesize\ttfamily,
  keywordstyle=\codekwstyle{},%
  columns=fullflexible,
 commentstyle=\color{mygreen},
  escapeinside={(*@}{@*)}
}[keywords,comments,strings]%

\newcommand{\metavar}[1]{$#1$}
\newcommand{\metavars}[1]{$\MakeLowercase{#1}$}
\newcommand{\binop}{\diamond}
\newcommand{\transformsto}{\big\Downarrow}

\lstdefinestyle{sql_style}{
morekeywords={WITH,OVER},
commentstyle=\itshape\color{red}, keywordstyle=\codekwstyle{},
  aboveskip=1pt,%\smallskipamount,
  belowskip=1pt,%\negsmallskipamount,
  lineskip=-0.2pt,
, basicstyle=\footnotesize\ttfamily
}

\title{PyTond: Efficient Python Data Science on the Shoulders of Databases}

\author{\IEEEauthorblockN{Hesam Shahrokhi}
\IEEEauthorblockA{\textit{School of Informatics} \\
\textit{University of Edinburgh}\\
Edinburgh, United Kingdom \\
hesam.shahrokhi@ed.ac.uk}
\and
\IEEEauthorblockN{Amirali Kaboli}
\IEEEauthorblockA{\textit{School of Informatics} \\
\textit{University of Edinburgh}\\
Edinburgh, United Kingdom \\
amirali.kaboli@ed.ac.uk}
\and
\IEEEauthorblockN{Mahdi Ghorbani}
\IEEEauthorblockA{\textit{School of Informatics} \\
\textit{University of Edinburgh}\\
Edinburgh, United Kingdom \\
mahdi.ghorbani@ed.ac.uk}
\and
\IEEEauthorblockN{Amir Shaikhha}
\IEEEauthorblockA{\textit{School of Informatics} \\
\textit{University of Edinburgh}\\
Edinburgh, United Kingdom \\
amir.shaikhha@ed.ac.uk}
}

\maketitle

\begin{abstract}
Python data science libraries such as Pandas and NumPy have recently gained immense popularity. Although these libraries are feature-rich and easy to use, their scalability limitations require more robust computational resources. In this paper, we present \system{}, an efficient approach to push the processing of data science workloads down into the database engines that are already known for their big data handling capabilities. Compared to the previous work, by introducing \xir{}, our approach can capture a more comprehensive set of workloads and data layouts. Moreover, by doing IR-level optimizations, we generate better SQL code that improves the query processing by the underlying database engine. Our evaluation results show promising performance improvement compared to Python and other alternatives for diverse data science workloads.
% \system{} exhibits a geometric average performance improvement of 39$\times$ compared to Python in the TPC-H benchmark and achieves a 4.5$\times$ speedup in our hybrid workloads.
\end{abstract}


\section{Introduction}\label{introduction}

Python, as a user-friendly language with extensive library support, has become a de-facto standard for data science. While Python excels in prototyping, its interpreted nature hinders its scaling on larger datasets and more complex workloads. This limitation also affects the performance of popular Python libraries like Pandas~\cite{pandas} and NumPy~\cite{numpy}. To overcome the Python interpretation overhead, these libraries have implemented their APIs in C language (using Python/C API) and utilize them as pre-compiled kernels. However, they still miss the chance to harness global program optimization opportunities such as fusion and dead-code elimination. Consequently, users still need to enhance their computational resources to achieve the expected performance, specifically when processing large amounts of data in data science pipelines.

In recent years, researchers have proposed innovative approaches to overcome the scalability constraints of Python and its associated libraries by leveraging techniques from compilers and database communities~\cite{blacher2022machine, blacher2023efficient, fischer2022snakes, hagedorn2021putting, petersohn2020towards, emani2024pyfroid, klabe2022accelerating, lam2015numba, palkar2018weld,  shahrokhi2023building, spiegelberg2021tuplex,Neumann11, dbtoaster, legobase_tods,dblablb, DBLP:journals/debu/ViglasBN14,DBLP:journals/jfp/ShaikhhaD018,DBLP:journals/pvldb/JungmairKG22}. One of the mainstream solutions~\cite{blacher2022machine, blacher2023efficient, fischer2022snakes, hagedorn2021putting, petersohn2020towards, emani2024pyfroid} focuses on leveraging the proven computational capabilities of database engines. These approaches translate the source code into SQL, allowing it to be executed on any standard relational database engine (portability). As data often resides inside databases, this approach can also bring the computation near the data and lessen unnecessary data movements. Moreover, by delegating query optimizations to query engines, the solution yields an efficient execution. 

Although transforming Python Data Science to SQL has promising benefits, the existing proposals have limitations. First, their translation coverage is restricted to a limited set of Pandas or NumPy APIs; For example, to the best of our knowledge, none of the existing proposals fully support the TPC-H benchmark~\cite{hagedorn2021putting}. Second, their generated code is not idiomatic SQL to fully exploit the potential performance benefits offered by the query engines. Lastly, the efforts on translating linear algebra expressions (e.g., NumPy APIs) to SQL are limited to the sparse data layout, which is not optimized for workloads with dense vectors/matrices~\cite{blacher2022machine,blacher2023efficient}.

This paper presents \system{}, an automated approach for in-database execution of Python data science workloads. \system{} takes workloads written in Pandas and/or NumPy and generates their semantically equivalent SQL code. The translated code is rewritten to idiomatic SQL code to make it more prone to optimizations offered by the underlying query processing engine. Finally, the translation process supports both sparse and dense data layouts for tensors (e.g., vectors and matrices), which enables \system{} to generate optimized SQL code for linear algebra operations.

\begin{table*}[!t]
    \caption{A summary of state-of-the-art in-database Python execution approaches. Blatcher et al.~\cite{blacher2023efficient} only supports the \code{einsum} API in NumPy on sparse data layout. Grizzly~\cite{hagedorn2021putting} has limited support for different Pandas APIs. PyTond provides support for data stored in dense and sparse layouts.}
    \centering
    % \scriptsize
    % \begin{scriptsize}
    \begin{tabular}{|l|c|c|c|c|c|}
    \hline
        \textbf{Approach} &
        \textbf{Generic Python} &
        \textbf{Pandas} &
        \textbf{NumPy} &
        \textbf{Multiple Data Layout} &
        \textbf{SQL Rewriting}
        \\ \hline \hline
        ByePy~\cite{fischer2022snakes} &
        \supfull &
        \supnone  &
        \supnone  &
        \suphalf &
        \supnone
        \\ \hline
        Blatcher et al.~\cite{blacher2023efficient} &
        \supnone &
        \supnone &
        \suphalf &
        \suphalf &
        \supnone
        \\ \hline
        Grizzly~\cite{hagedorn2021putting} &
        \suphalf &
        \suphalf &
        \supnone &
        \suphalf &
        \supnone
        \\ \hline
        PyFroid~\cite{emani2024pyfroid} &
        \supnone &
        \supfull &
        \supnone &
        \suphalf &
        \suphalf
        \\ \hline
        \textit{\textbf{PyTond} (This Approach)} &
        \supnone &
        \supfull &
        \supfull &
        \supfull &
        \supfull
        \\ \hline
    \end{tabular}
    \vspace{-0.25cm}
    \label{tab-approches-summary}
\end{table*}

The contributions of this paper are as follows:
\begin{itemize}
    \item We present \system{}, a framework for the portable in-database execution of hybrid Python data science workloads. This is achieved by translating Pandas/NumPy APIs to SQL code. We discuss its high-level design that offers expressiveness and efficiency and the background in Section~\ref{backgroundoverview}.
    
    \item We show that \system{} is unique in capturing Pandas (relational algebra), NumPy (linear algebra), and their hybrid workloads while considering different data layouts. The key enabling technology in achieving expressiveness is \xir{}, a novel intermediate language that enables an extensible translation of Pandas/NumPy APIs (Section~\ref{expressiveness}). 
    
    \item We apply optimizations on the \xir{} program to generate SQL code that paves the way to further optimizations and more efficient processing by the underlying database engine. Our discussion on the efficiency of the approach is covered in Section~\ref{efficiency}.
       
    \item We conduct benchmarks (Section~\ref{evaluation}) to showcase the superiority of our approach over Python and state-of-the-art for various workloads, including all TPC-H queries and hybrid data science workloads.
    
    % Compared to Python, using 4 threads, our approach is highly efficient with a geometric average performance improvement of 39$\times$ in the TPC-H benchmark and 4.5$\times$ for our hybrid workloads. On a single thread, \system{} still shows a geometric average of 16$\times$ and 1.7$\times$ speedup for the same set of experiments, respectively.
\end{itemize}

Table~\ref{tab-approches-summary} summarizes the characteristics of the existing research on in-database scalable Python data analytics and compares them with \system{}. We refer to Section~\ref{related} for a more detailed discussion on state-of-the-art.
\section{Background and Overview}\label{backgroundoverview}
In this section, we provide the background knowledge that is necessary for a better understanding of the work. Then, we present the design of our approach. 

\subsection{Background}\label{background}
In this section, we provide contextual information and present the high-level design of our approach. 

\smartpara{Background on Pandas}
The Pandas library is a popular tool for relational algebra (RA) operations in Python. Pandas introduces \textit{DataFrame} as its primary data structure. The DataFrame data structure is a flexible representation of 2-dimensional data; it is easy to add, modify, and remove the columns and rows of a given DataFrame by calling their related APIs. Table~\ref{tab-pandas-summary} shows a set of primary Pandas APIs.

The \codekwpy{merge} API works similarly to a join in RA. The \code{how} argument of this API determines the type of requested operations including the Cartesian product, inner join (default value), left, right, and full outer join.

To demonstrate the \codekwpy{pivot\_table} API, consider the following expression:

\begin{lstlisting}
df.pivot_table(index='a', columns='b', values='c', func='sum')  
\end{lstlisting}

\begin{table}[!t]
    \caption{A summary of primary Pandas APIs.}
    \centering
    % \scriptsize
    % \begin{scriptsize}
    \begin{tabular}{|l|l|}
    \hline
        \textbf{API} &
        \textbf{Description}
        \\ \hline \hline
        \begin{lstlisting}
df[col] | df.col
        \end{lstlisting} &
        Column selection
        \\ \hline
        \begin{lstlisting}
df[condition]
        \end{lstlisting} &
        Rows filtering
        \\ \hline
        \begin{lstlisting}
df.head(n)
        \end{lstlisting} &
        Top \code{n} rows selection
        \\ \hline
        \begin{lstlisting}
df.col.unique()
        \end{lstlisting} &
        Distinct column values
        \\ \hline
        \begin{lstlisting}
df.sort_values(by, ascending)
        \end{lstlisting} &
        Sorting on columns list 
        \\ \hline
        \begin{lstlisting}
df.apply(func)
        \end{lstlisting} &
        Mapping using \code{func}
        \\ \hline
        \begin{lstlisting}
df.aggregate(func)
        \end{lstlisting} &
        Reduction using \code{func} 
        \\ \hline
        \begin{lstlisting}
df.groupby(by, axis)
        \end{lstlisting} &
        Grouping
        \\ \hline
        \begin{lstlisting}
df1.merge(df2, how, on)
        \end{lstlisting} &
        Joining DataFrames
        \\ \hline
        \begin{lstlisting}
df1.isin(df2)
        \end{lstlisting} &
        Containment filtering
        \\ \hline
        \begin{lstlisting}
df.pivot_table(index,...,func)
        \end{lstlisting} &
        Making a pivot table
        \\ \hline
    \end{tabular}
    \label{tab-pandas-summary}
\end{table}

In this example, the output DataFrame has a column \code{a}, and its distinct values are used as group representatives. Then, for all rows that fall into each group, a column will be associated with each distinct value of the column \code{b}. In any of those columns, we have the sum of values in column \code{c}. A sample input DataFrame (\code{df}) and the result of applying the above API call on it is shown below:

\vspace{1em}
\begin{center}
\scriptsize
\begin{tabular}{c c c}
    \begin{tabular}{|c|c|c|}
    \hline
        \textbf{a} &
        \textbf{b} &
        \textbf{c}
        \\ \hline
        \hline x & v1 & 10 \\
        \hline y & v3 & 30 \\
        \hline y & v1 & 60 \\
        \hline z & v2 & 20 \\
        \hline y & v3 & 40 \\
        \hline x & v2 & 60 \\
        \hline z & v2 & 50 \\
        \hline
    \end{tabular}&
$\longrightarrow$&
    \begin{tabular}{|c|c|c|c|}
    \hline
        \textbf{a} &
        \textbf{v1} &
        \textbf{v2} &
        \textbf{v3}
        \\ \hline
        \hline x & 10 & 60 & 0 \\
        \hline y & 60 & 0 & 30 \\
        \hline z & 0 & 70 & 0 \\   
        \hline  
    \end{tabular}
\end{tabular}
\end{center}
\vspace{1em}

\smartpara{Background on NumPy}
The NumPy library is a popular tool for linear algebra (LA) computations in Python. This library is based on fine-tuned kernels that offer an efficient execution. The basic data structure in NumPy is called \textit{array} which is an efficient representation of a tensor of an arbitrary order. Here is an example of tensors of different orders defined in NumPy:

\begin{lstlisting}
# Order-1 Tensor (Vector)
vector = np.array([1, 2, 3])
# Order-2 Tensor (Matrix)
matrix = np.array([[1, 2, 3], [4, 5, 6], [7, 8, 9]])
# Order-3 Tensor (3D Tensor)
tensor_3d = np.array([[[1, 2, 3], [4, 5, 6]], [[7, 8, 9], [10, 11, 12]]])
\end{lstlisting}

One of the interesting and powerful APIs NumPy provides is called \codekwpy{einsum}. Adopted from the concept of Einstein Notation (Summation)~\cite{einot}, this API provides a concise, comprehensive, and efficient way to write linear algebra operations in Python. Each \codekwpy{einsum} API call takes a list of arguments. The first argument is a string that defines the requested LA operation, while the rest are the required operands. Table~\ref{tab-einsum-summary} lists some common LA operations and their equivalent NumPy expression. As it can be inferred from the table, many of the specific APIs for LA in NumPy (e.g., \codekwpy{sum}, \codekwpy{transpose}, \codekwpy{inner}, \codekwpy{matmul}) can be expressed in terms of an \codekwpy{einsum} API. However, APIs such as \codekwpy{all}, \codekwpy{nonzeros}, \codekwpy{round},  and \codekwpy{compress} cannot be expressed solely by using \codekwpy{einsum}.

\begin{table}[t]
    \caption{A list of common LA operations in NumPy and their equivalent \code{einsum} expressions.}
    \centering
    \begin{tabular}{|l|l|l|}
    \hline
        \textbf{NumPy API} &
        \textbf{Args of \codetiny{np.einsum}} &
        \textbf{Description}
        \\ \hline \hline
        \begin{lstlisting}
m.sum(axis=0)
        \end{lstlisting} &
        \codetiny{('ij->j',m)} &
        Matrix ColSum
        \\ \hline  
        \begin{lstlisting}
m.sum(axis=1)
        \end{lstlisting} &
        \codetiny{('ij->i',m)} &
        Matrix RowSum
        \\ \hline  
        \begin{lstlisting}
m.sum()
        \end{lstlisting} &
        \codetiny{('ij->',m)} &
        Matrix Sum
        \\ \hline  
        \begin{lstlisting}
inner(v1, v2)
        \end{lstlisting} &
        \codetiny{('i,i->',v1,v2)}
        &
        Vector Inner Prod.
        \\ \hline  
        \begin{lstlisting}
outer(v1, v2)
        \end{lstlisting} &
        \codetiny{('i,j->ij',v1,v2)} &
        Vector Outer Prod.
        \\ \hline
        \begin{lstlisting}
m.transpose()
        \end{lstlisting} &
        \codetiny{('ij->ji',m)} &
        Matrix Transpose
        \\ \hline  
        \begin{lstlisting}
matmul(m1,m2)
        \end{lstlisting} &
        \codetiny{('ij,jk->ik',m1,m2)} &
        Matrix Mult.
        \\ \hline  
        \begin{lstlisting}
v.all()
        \end{lstlisting} &
        \codetiny{-} &
        All Set Check
        \\ \hline
        \begin{lstlisting}
v.nonzero()
        \end{lstlisting} &
        \codetiny{-} &
        Non-Zeros Index
        \\ \hline
        \begin{lstlisting}
v.round()
        \end{lstlisting} &
        \codetiny{-} &
        Rounding
        \\ \hline
        \begin{lstlisting}
v.compress(c)
        \end{lstlisting} &
        \codetiny{-} &
        Slicing
        \\ \hline        
    \end{tabular}
    \label{tab-einsum-summary}
\end{table}

\subsection{Overview of Approach}
\begin{figure}[t]
\centering
\includegraphics[width=0.45\textwidth]{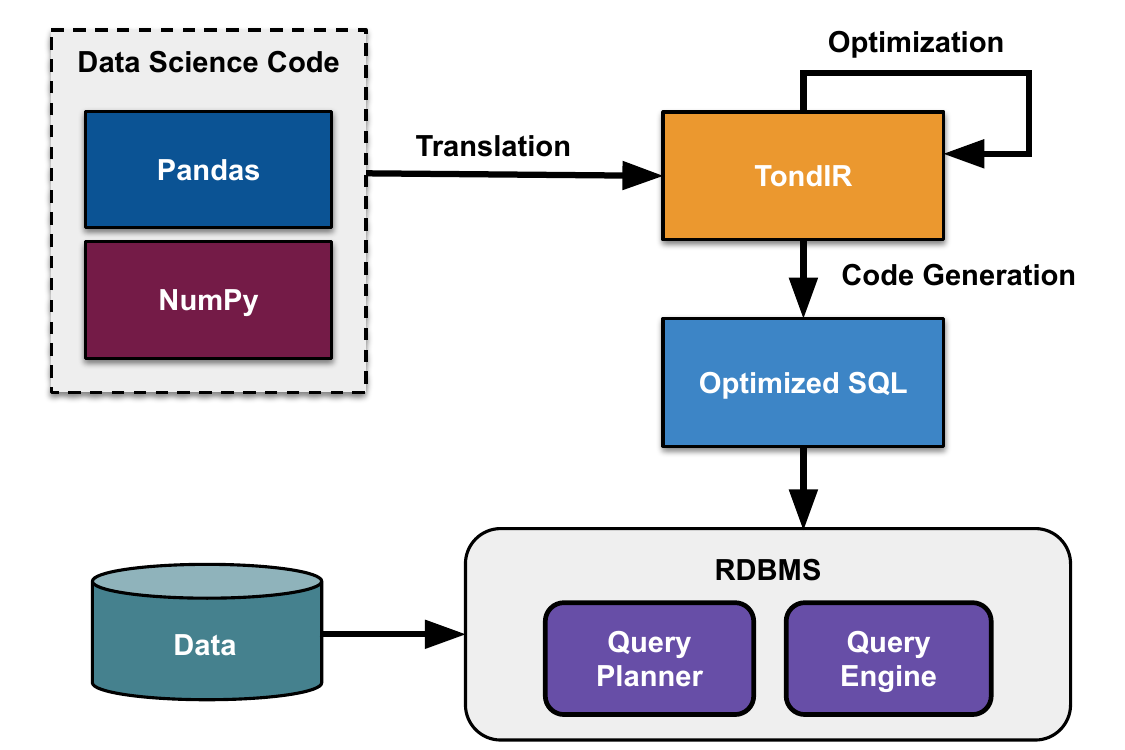}
\caption{High-level design of \system{}.}
\label{fig:arch}
\end{figure}

The high-level overview of \system{} is depicted in Figure~\ref{fig:arch}. Here we elaborate on each element of the presented design.

\smartpara{Data Science Code}
\system{} supports the data science workloads that are written using Pandas, NumPy, or a combination of them. Since our approach is based on static analysis of the source code, data scientists do not need to change their program logic or even the imported libraries. The only required change is to add the \decorator{} decorator on top of their functions (cf. Section~\ref{expressiveness}).

\smartpara{Translation}
The translator module starts by finding the decorated functions in the source code and constructs the abstract syntax tree (AST) of them. Then, it transforms the ASTs into our tailored intermediate representation (IR), based on pre-defined translation rules. The translation process and its various challenges will be discussed in Sections~\ref{pythontoir}, \ref{pandastoir}, and \ref{numpytoir}. 

\smartpara{\xir{}}
\xir{} is our gateway to expressiveness and our lever to improve efficiency. This IR is designed to capture the intended workloads and is amenable to applying a variety of optimizations. An in-depth analysis of this IR will be done in Section~\ref{irdesign}.

\smartpara{Optimization}
The optimizer module gets an \xir{} code and applies a set of well-defined optimization rules to construct a more efficient version of the input. In Section~\ref{efficiency}, we will investigate the transformations done during the optimization phase. We will also assess its effects on the overall performance of \system{} (Section~\ref{evaluation}).  

\smartpara{Code Generation}
After the optimization is done, we translate the optimized \xir{} code to standard SQL. Our SQL code generation challenges are discussed in Section~\ref{irtosql}.

\smartpara{Data}
The data that is fed into the data science pipelines is usually stored either in files (e.g. CSV files) or databases. For huge data sizes, the poor performance of Pandas popular file reader API (\codekwpy{read\_csv}) pushes the developers towards using the databases as their main data storage locations. The capability of databases for handling massive data besides their recent advances (e.g. in-memory databases~\cite{kersten2018everything}) also justifies this preference. By taking the opportunity of data being kept in databases, our approach can gain access to the data and bring the computation (execution of the optimized SQL code) near the data. Before the execution, by querying the database catalog, our approach can access the meta-data information (e.g., schema and integrity constraints) and use it during the optimization process. We discuss the potential benefits of having such knowledge in Section~\ref{expressiveness}. 

Focusing on the numerical data, specifically matrices, regardless of the storage location, the data is usually kept in a 2-dimensional format. We refer to it as a \textit{dense layout},\footnote{Blatcher et al.~\cite{blacher2022machine} call this representation \textit{database-friendly}.} comparing it to \textit{sparse layouts}. To construct a sparse (COO) representation of a given dense input, we first create an ID column (starting from 0), an incremental integer column associated with each row. Then, we virtually assign a unique integer number (starting from 0) to each column. Finally, we transform the given matrix to a new 2-dimensional representation with this header: (row\_ID, col\_ID, val). This layout is beneficial whenever the input data is sparse (many values are None or 0) but needs extra transformations on input/output.

In \system{}, we provide the flexibility of using both dense and sparse layouts by passing this information to function decorators (Section~\ref{expressiveness}). Having the support for both layouts, specifically for the LA workloads, makes it possible to process numerical data in its natural format (dense) while benefiting from the advantages of sparse layout when the data is sparse. The state-of-the-art~\cite{blacher2023efficient} only supports the sparse format.

\smartpara{RDBMS}
As the final step in our pipeline, the optimized SQL code is passed to the underlying relational database management system (RBDMS) for further optimization and final execution. The RDBMS receives the query, does a bunch of thoughtful optimizations over it, and passes it to its query engine to be executed on the already-loaded data. It is worth mentioning that the optimizations that are done in both \xir{} and query planner level, could not be easily done by the developers or in the interpreted Python environment.
Last but not least, the query engine executes the generated physical plan of the query in a fine-tuned hardware-conscious fashion. This execution can exploit the recent advances in query processing including push-based and vectorized engines~\cite{kersten2018everything}.  
\section{\xir{}}\label{expressiveness}

\begin{table*}[t!]
    \caption{Grammar of \xir{}. In atoms we have $\theta \in \{$ \code{<,<=,=,<>,>=,>} $\}$}
    \label{tbl:grammar}
    \centering
    \begin{tabular}{|r r c l|l|}
      \hline
        Program & \metavar{P} & $::=$ & \metavar{R} $\mid$ \metavar{P} \metavar{R} &  List of rules. \\ \hline
        Rule & \metavar{R} & $::=$ & \metavar{H} \code{:-} \metavar{B}\code{.} & Head (access) and body (products). \\ \hline
        Head & \metavar{H} & $::=$ & \metavars{r}[\codekw{group}\code{(}$\overline{\text{\metavars{X}}}$\code{)}][\codekw{sort}\code{(}$\overline{\text{\metavars{X}}}, \overline{\text{\metavars{B}}}$\code{)}[\codekw{limit}\code{(}$n$\code{)}]] 
        & Aggregates over a relation. Arguments in [] are optional.
        \\ \hline
        Relation & \metavars{r} & $::=$ & \metavar{X}\texttt{(}$\overline{x}$\texttt{)}
        & Access to a relation \metavar{X} with variables $\overline{x}$.
        \\ \hline
        Body & \metavar{B} & $::=$ & \metavars{A}  $\mid$ 
        \metavar{B}  $,$  \metavars{A}
        & Body is a chain of atoms.
        \\ \hline
        Atom & \metavars{A} & $::=$ & \metavars{r} $\mid$ \code{[$\overline{\code{<c>}}$] $\mid$ \codekw{exists}\code{(}\metavar{B}\code{)} $\mid$ \metavars{X} $\theta$ \metavars{T}}
        & Relation access, constant relation, existential filter, or logical/assignment.\\ \hline
        Term & \metavars{T} & $::=$ & 
        \metavars{x} $\mid$ \metavars{agg(t)} $\mid$ \metavars{ext(\overline{x})} $\mid$ \codekw{if}\code{(}$t, t, t$\code{)}
        $\mid$ \metavars{T} $\binop$ \metavars{T} $\mid$ \metavars{c} & Variable, aggregation, external function, conditional, binary op, or constant. \\ \hline
        Constants & \metavars{c} & $::=$  &  
        \metavars{N} $\mid$ \metavars{F} $\mid$ \metavars{B} $\mid$ \metavars{S} & integer, floating point, boolean, or string.
        \\ \hline
    \end{tabular}

\end{table*}

In this section, we present \xir{}, our intermediate representation designed to capture Python data science workloads efficiently and pave the way for subsequent optimizations. We start by specification of the \xir{} design. Following that, we explain the integration of our translation approach into Python that ensures a seamless transition to \xir{}. Next, in dedicated sections, we discuss the translation process from Pandas and NumPy to \xir{}. Then, we elaborate on our strategy for translating \xir{} to SQL. Ultimately, we wrap up all components and demonstrate an end-to-end translation from Python to SQL.

\subsection{\xir{} Design}\label{irdesign}
To design an appropriate intermediate representation (IR) for our approach, we consider its ability to express hybrid data science workloads involving relational and linear algebra, and its flexibility for applying different optimizations. To this end, we designed an IR inspired from Datalog-like logic programming languages~\cite{DBLP:books/aw/AbiteboulHV95,DBLP:journals/pacmmod/ShaikhhaSSN24,DBLP:journals/pacmpl/GhorbaniHHS23,abo2016faq,rel} with the mentioned properties. The grammar of this IR is shown in Table~\ref{tbl:grammar}.

\smartpara{Grammar} Each program (\metavar{P}) is composed of several rules (\metavar{R}). Each rule is in the form of an assignment from a body (\metavar{B}) to a head (\metavar{H}). A head is an access to a relation (\metavars{r}) with an optional group-by ([\codekw{group}\code{(}$\overline{\text{\metavars{X}}}$\code{)}]) and an optional ascending/descending ($\overline{\text{\metavars{B}}}$) sort with/without limiting ([\codekw{sort}\code{(}$\overline{\text{\metavars{X}}}, \overline{\text{\metavars{B}}}$\code{)}[\codekw{limit}\code{(}$n$\code{)}]]), over a list of variables. The relation access (\metavars{r}) is described by providing the relation name (\metavar{X}) and a list of variables assigned to each relation column ($\overline{x}$). The relation column names are bound to the position of each variable \metavar{X} inside $\overline{x}$. This will guarantee sound code generation for column names even after local changes during the translation and optimization. The body is a chain of atoms (\metavars{a}). Each atom can be a relation access, creation of a constant relation, an existential filter (\codekw{exists}\code{(}\metavar{B}\code{)}), or a comparison/assignment ($\theta$) between a variable (\metavars{X}) and a term (\metavars{T}). If the left side of an assignment is an already defined variable, we consider the operation an equality comparison rather than an assignment.
Finally, a term is an aggregation over another term (\metavars{agg(x)}), an external function call over variables (\metavars{ext(\overline{x})}), a conditional statement (\codekw{if}\code{(}$t, t, t$\code{)}), a binary operation over terms (\metavars{T} $\binop$ \metavars{T}) or a constant value (integer (\metavars{n}), floating point (\metavars{f}), string (\metavars{s}), or boolean (\metavars{b})). Binary operations include arithmetic, and/or, like, etc.  

\smartpara{Contextual Information} Collection of contextual information about the workload can guide the optimizations on the generated \xir{}. In \system{}, we use two distinct sources to gather such information: database catalog and decorator arguments. 

Since we assume that the data is imported into the database, \system{} queries the DBMS catalog to implicitly collect contextual information about the dataset, and embed them into \xir{} constructs. The extracted information includes but is not limited to the table constraints (e.g., primary key, foreign key, uniqueness), and schema information (e.g. cardinality, column names and types).

The other source of collecting contextual information is through the function decorators. As discussed in Section~\ref{backgroundoverview}, to transform a Python function to SQL using \system{}, users need to add \decorator{} decorator on top of their functions. The contextual information can be explicitly passed to \system{} as the arguments of function decorators. 

\subsection{Python Embedding}\label{pythontoir}
\system{} scans the Python code to identify functions decorated with \decorator{}. The abstract syntax tree (AST) is extracted for each function using the \code{ast} module. Then, the extracted AST is passed to the translation preprocessing pipeline.

\smartpara{Normalization} \system{} converts the AST to A-Normal Form (ANF)~\cite{DBLP:conf/pldi/MaurerDAJ17,DBLP:conf/pldi/FlanaganSDF93}. In the ANF representation, we extract each nested logic, assign it to a variable, and put a variable access instead of the nested logic. This makes the translation easier as we only need to define rules for simple expressions. Here is an example of converting a Pandas code to ANF:

\vspace{1em}
\begin{lstlisting}
res = (df1[df1.b>10]['a']).merge((df2[df2.y=='r']['x']), left_on='a', right_on='x')
\end{lstlisting}
\begin{center}$\transformsto$\end{center}
\begin{lstlisting}
v1 = df1.b>10
v2 = df1[v1]
v3 = v2['a']
v4 = df2.y=='r'
v5 = df2[v4]
v6 = v5['x']
res = v3.merge(v6, left_on='a', right_on='x')
\end{lstlisting}
\vspace{1em}

As shown above, the nested projection and filtering logic are decoupled into consecutive assignments to globally unique variables. After this program transformation, \system{} only needs to use a single transformation rule to transform each line of source code. In the normalization phase, to preserve the program semantics, the input variable names (\code{df1} and \code{df2}) remain unchanged.

\begin{table*}[!t]
    \caption{Translation of sample Pandas/NumPy APIs to \xir{}. \code{df}, \code{df1}, \code{df2}, \code{m}, \code{m1}, \code{m2} all have \code{n} columns. The result of each API call is stored in variable \code{R}. For example, the first Python code equals to: \code{R=df[col]}}
    \centering
    \begin{tabular}{|c|l|l|}
    \hline
        \textbf{API Library} &
        \textbf{Python} &
        \textbf{\xir{}}
        \\ \hline
        \hline
        Pandas &
        \begin{lstlisting}
df[col]
        \end{lstlisting} &
        \begin{lstlisting}[language=TondIR]
R(col) :- df(c1, ... , cn).
        \end{lstlisting} \\ \hline
        Pandas &
        \begin{lstlisting}
df[condition]
        \end{lstlisting} &
        \begin{lstlisting}[language=TondIR]
R(c1, ... , cn) :- df(c1, ... , cn), (condition).
        \end{lstlisting} \\ \hline
        Pandas &
        \begin{lstlisting}
df.aggregate(func)
        \end{lstlisting} &
        \begin{lstlisting}[language=TondIR]
R(c1, ... , cn) :- df(c1, ... , cn),
                 (c1=func(c1)), ... , (cn=func(cn)).
        \end{lstlisting} \\ \hline
        Pandas &
        \begin{lstlisting}
df1.merge(df2, 
          how='inner', 
          on='cx')
        \end{lstlisting} &
        \begin{lstlisting}[language=TondIR]
R(x1, ... , x(2n-1)) :- df(a1, ... , x , ... , an),
                  df(b1, ... , cx , ... , bn).
        \end{lstlisting} \\ \hline
        Pandas &
        \begin{lstlisting}
df.sort_values(by, asc).head(n)
        \end{lstlisting} &
        \begin{lstlisting}[language=TondIR]
R(c1, ... , cn) sort(by, ascending) limit(n) :-
                df(c1, ... , cn).
        \end{lstlisting} \\ \hline
        Pandas &
        \begin{lstlisting}
df.groupby(col).sum()
        \end{lstlisting} &
        \begin{lstlisting}[language=TondIR]
R(c1, ... , cn) group([col]) :- df(c1, ... , cn),
                (c1=sum(c1), .... , cn=sum(cn)).
        \end{lstlisting} \\ \hline
        NumPy &
        \begin{lstlisting}
v.all(axis=1)
        \end{lstlisting} &
        \begin{lstlisting}[language=TondIR]
R(c1) :- v(ID, c1), (c1=min(c1)).
        \end{lstlisting} \\ \hline
        NumPy &
        \begin{lstlisting}
v.nonzero(axis=1)
        \end{lstlisting} &
        \begin{lstlisting}[language=TondIR]
R(ID) :- v(ID, c1), (c1!=0).
        \end{lstlisting} \\ \hline
        NumPy &
        \begin{lstlisting}
v.round()
        \end{lstlisting} &
        \begin{lstlisting}[language=TondIR]
R(ID, c1) :- v(ID, c1), (c1=round(c1)).
        \end{lstlisting} \\ \hline
        NumPy &
        \begin{lstlisting}
m.compress(mask, axis=1)
        \end{lstlisting} &
        \begin{lstlisting}[language=TondIR]
R(ID, cx, ... , cy) :- m(ID, c1, ... , cn)).
        \end{lstlisting} \\ \hline
        NumPy &
        \begin{lstlisting}
einsum('ii->i',m)
        \end{lstlisting} &
        \begin{lstlisting}[language=TondIR]
R(ID, c1) :- m(ID, c1, ... , cn),
(c1=if(ID=1, c1, if(ID=2, c2, ... , if(ID=n, cn) ... ))) 
        \end{lstlisting} \\ \hline
        NumPy &
        \begin{lstlisting}
einsum('ij,ij->ij', m1, m2)
        \end{lstlisting} &
        \begin{lstlisting}[language=TondIR]
R(ID, c1, ... , cn) :- m1(ID, a1, ... , an),
                               m2(ID, b1, ... , bn),
                               (c1=a1*b1), ... , (cn=an*bn).
        \end{lstlisting} \\ \hline
        
        \hline
    \end{tabular}
    \label{tbl:all-translations-summary}
\end{table*}

\smartpara{Type Inference} \system{} fetches the contextual information (cf. Section~\ref{irdesign}) from decorator arguments and database catalog. Then, having the normalized program and its context, \system{} will do type inference. The extracted type information will be used in translation/optimization pipelines.

\smartpara{Relation Access Renaming} To avoid variable name collisions in \xir{} translation/optimization, \system{} assigns a unique name to each variable in each body relation access (\code{r}).

\subsection{Pandas to \xir{}}\label{pandastoir}

The \xir{} equivalents of sample Pandas APIs are listed in Table~\ref{tbl:all-translations-summary}. Due to the flexible and expressive nature of \xir{}, the common Pandas expressions can be easily transformed to this IR. 

To translate the \codekwpy{merge} API, \system{} takes different translation approaches based on the \code{how} attribute of the API. If it is a Cartesian product (\code{how='cross'}), \system{} assigns unique names for all variables in both relations. If it is an inner join (\code{how='inner'}), \system{} assigns unique names to all variables except the join variables, which will have similar names. For the case of full, left, and right outer joins, it takes an approach similar to the Cartesian. However, it introduces special external atoms \metavars{ext(\overline{x})} that contain information about the join type and joined columns. These atoms are called \metavars{outer\_full(\overline{x})}, \metavars{outer\_left(\overline{x})}, and \metavars{outer\_right(\overline{x})}.

Although the presented APIs are only a subset of what is covered by \system{} and Pandas, the extensibility of \xir{} makes the coverage of the other APIs feasible. Next, we focus on the challenges in translation from Pandas to \xir{}.

\smartpara{Implicit Renaming}
In Pandas, when we use the \codekwpy{merge} API to combine DataFrames with overlapping column names, the API will automatically rename the shared columns of the first DataFrame to \code{[col\_name]\_x} and their corresponding columns in the second DataFrame to \code{[col\_name]\_y}. The \code{\_x} and \code{\_y} constants can be changed by setting the \code{suffixes} argument of \codekwpy{merge} API. If the shared column is the one used for the merging operation, the API retains the original column name and includes only one instance of it in the merged result. An example of a \codekwpy{merge} operation with the mentioned properties is shown below:

\begin{lstlisting}
df1 = ... # df1 column names: [a,b,c]
df2 = ... # df2 column names: [a,c,d]
df3 = df1.merge(df2, left_on='a', right_on='a') 
\end{lstlisting}

In this example, the header of the output DataFrame (\code{df3}) will be [a, b, c\_x, c\_y, d]. During its type inference phase, \system{} considers this implicit renaming to generate correct types for \codekwpy{merge} API calls.

\smartpara{Implicit Joins} Pandas allows users to append to the columns of a DataFrame. This can be expressed as an implicit join operation.
Consider the following example in Pandas:

\begin{lstlisting}
df1 = ... # df1 column names: [a,x,y]
df2 = ... # df2 column names: [b,z]
df3 = DataFrame()
df3['a'] = df1['a']
df3['b'] = df2['b']
\end{lstlisting}

In this example, we create a new column on the left-hand side DataFrame by using the values from a column on the right-hand side. Note that the program can be even more complex when we update (instead of create) a column in \code{df3} or have more complex expressions from different DataFrames on the right-hand side. The high abstraction level of DataFrames in Pandas can hide the challenges of handling such scenarios. However, to generate SQL, we must explicitly express the implicit operations done at Pandas background. Back to our example, since \code{df3} is initially empty, we can generate a projection from \code{df1} that only projects column \code{a}. But, in the next line, we need to add a column \code{b} from another DataFrame (\code{df2}) and it needs a join between \code{df3} and \code{df2}. Here, there is an assumption that the source and target columns have equal rows. \system{} generates the following \xir{} in this scenario:

\vspace{1em}
\begin{lstlisting}[language=tondir]
df3(a) :- df1(a, x, y).
df3(ID, a) :- df3(a), (ID=UID()).
df2(ID, b, z) :- df2(b, z), (ID=UID()).
df3(a, b) :- df3(ID, a), df2(ID, b, z).
\end{lstlisting}
\vspace{1em}

The \code{UID} term is an external atom (\metavars{ext(\overline{x})}) and is used to create an identity (PK) column for a relation. In this code, \system{} automatically adds the necessary rules/terms to make the implicit join on \code{df2} and \code{df3} explicit. As it will be discussed in \ref{efficiency}, \system{} optimizer will finally remove unnecessarily introduced rules/terms (e.g. when there is a self-join).

\smartpara{Pivot Translation}
We presented the interface of Pandas \code{pivot\_table} API in Table~\ref{tab-pandas-summary}. To translate this API, we need information about the distinct values specified by the \code{columns} argument. This information can be provided via the decorator or by querying the target columns before code generation. Considering the example from Section~\ref{background}, its equivalent \xir{} code is as follows:

\vspace{1em}
\begin{lstlisting}[language=tondir]
R1(a, v1, v2, v3) group(a) :- 
  R(a, b, c), (v1=sum(if(b=v1,c,0))),
  (v2=sum(if(b=v2,c,0))), (v3=sum(if(b=v3,c,0))).
\end{lstlisting}
\vspace{1em}

We must generate distinct columns for unique values in column \code{b}. To achieve this, we examine the value of this column in every row and determine to which designated columns (v1, v2, and v3) that particular value should contribute. Subsequently, we aggregate the contributed values for each column using a \codekw{sum} expression and assign the result to the respective column name. Finally, the embedding of the \code{group} clause ensures that accurate grouping is done within the program. It is worth noting that the information about the distinct values of \code{b} can be passed to \system{} using the \code{@pytond} decorator arguments. 

\subsection{NumPy to \xir{}}\label{numpytoir}
Table~\ref{tbl:all-translations-summary} also shows translations of a subset of NumPy APIs to \xir{}. In all NumPy APIs that use arrays in their input, we assume a unique \code{id} column stored in the input. We also generate/project such a column after each API call if the result is an array. The \codekwpy{all} API checks if all values in the vector are \codekwpy{True}. We implemented it by applying the \codekwpy{min} function to vector values. The \codekwpy{compress} API with \code{axis=1}, slices the array vertically (picks the columns that are selected by \code{mask}). The following section will cover the detailed discussion on \codekwpy{einsum} API translations.

\smartpara{Einsum Translation}
The \codekwpy{einsum} API translation in our approach has a two-phase pipeline. First, we compute an execution plan for a given \codekwpy{einsum} expression, and then we generate code by considering the extracted plan.

In the planning phase, we first take the information about data layout (dense vs. sparse) from the function decorator. If nothing is passed, we consider the default data layout, which is dense. If the data layout is sparse, we take the approach proposed by Blacher et al.~\cite{blacher2023efficient} but generate \xir{} (not SQL). The fully denormalized nature of the COO sparse layout allows us to easily generate \xir{} code for different \codekwpy{einsum} expressions. On the other hand, if the dense format is requested, we will continue planning for the given \codekwpy{einsum} expression as will be discussed. 

After thoroughly exploring possible binary \codekwpy{einsum} expressions (with matrix/vector operands), we came up with a minimal set of kernels listed in Table~\ref{tbl:einsum-kernels}. Our planner will use these kernels to construct the other possible \codekwpy{einsum} expressions. Here, there is a trade-off between automation and efficiency. We can implement more efficient kernels manually, but it needs much more effort. We decided to go with this minimal set of kernels and let the optimizer take the responsibility for optimizing the generated \xir{}. The translations for the ES3 and ES7 kernels are shown in the last two rows of Table~\ref{tbl:all-translations-summary}. ES7 has a final matrix reshaping part that is not shown in the table for the sake of brevity. However, a sample of such reshape will be presented in our final end-to-end example (Figure~\ref{end2end}). 

Now, we explain the translation of \codekwpy{einsum} expressions to our fundamental kernels based on an example binary \codekwpy{einsum} expression. Assume that we have the \codekwpy{einsum} expression \code{'ab,cc->ba'} as input.  This expression can be represented as \code{'ij,kk->ji'} in \xir{} since \code{a}, \code{b}, and \code{c} appeared in the first, second, and third non-repeated position of the expression respectively. Since variable \code{k} does not appear on the right-hand side, \xir{} needs to sum over the whole matrix (\code{'kk->'}). Therefore, \xir{} applies kernel ES3 over the right-hand side of the compression and transforms \code{'kk->k'}. Then, \xir{} proceeds to apply kernel ES1 on the same matrix to produce the expected result \code{'k->'}. Now, the computation is simplified and turned into \code{'ij,->ji'}. However, this computation is not a fundamental kernel yet. As the next step, \xir{} swaps the left-hand side and right-hand side of this computation and converts it to \code{',ij->ji'}. As the final step, \xir{} transposes the input using kernel ES4 and creates the \codekwpy{einsum} expression \code{',ji->ji'} which is kernel ES6. By using this tail of einsum operations (plan), \system{} generates appropriate \xir{} code for the given einsum expression. During the code generation for einsums, if either inputs to the \codekwpy{einsum} API are constant, \system{} will inline constant values inside its predefined kernels (constant folding).   

\begin{table}[t]
    \caption{List of base \codekwpy{einsum} kernels in \system{}. All other \codekwpy{einsum} expressions can be reduced to these kernels. Given an \codekwpy{einsum} expression, the characters \code{i}, \code{j}, and \code{k} show the first, second, and third non-repeated variables respectively. \code{\_} represents an scalar value. The kernels are presented as args of a NumPy \codekwpy{einsum}.}
    \centering
    \begin{tabular}{|c|l|l|}
        \hline
        \textbf{ID} & \multicolumn{1}{c|}{\textbf{Fundamental Kernels}} & \textbf{Description} \\
        \hline \hline
        ES1 & \code{('i->',v)} & Vector ColSum \\
        \hline  
        ES2 & \code{('ij->i',m)} & Matrix ColSum \\
        \hline  
        ES3 & \code{('ii->i',m)} & Diagonal to Column \\
        \hline
        ES4 & \code{('ij->ji',m)} & Transpose \\
        \hline
        ES5 & \code{(',->',s1,s2)} & Scalar Product \\
        \hline
        ES6 & \code{(',ij->ij',s1,m1)} & Scalar Times Matrix \\
        \hline
        ES7 & \code{('ij,ij->ij',m1,m2)} & Hadamard Product \\
        \hline
        ES8 & \code{('ij,ik->jk',m1,m2)} & Batch Vector Outer Product \\
        \hline
        ES9 & \code{('ij,ik->ij',m1,m2)} & Matrix-Vector Multiplication \\
        \hline
    \end{tabular}
    \label{tbl:einsum-kernels}
\end{table}

\xir{} also supports \codekwpy{einsum} expressions with more than two inputs (non-binary) using the Python \textit{opt\_einsum} library~\cite{opt_einsum}. First, \system{} passes the \codekwpy{einsum} expression to the opt\_einsum path optimizer. Then, opt\_einsum breaks the computation to several \codekwpy{einsum} computations with less number of inputs. Finally, \xir{} processes and translates the computations if all the \codekwpy{einsum} expressions produced by opt\_einsum are in a binary format. 

\subsection{\xir{} to SQL}\label{irtosql}
To generate SQL from \xir{}, we take each rule and create a SQL \codekw{WITH} clause that contains the SQL equivalent of the logic defined by the rule. Here is an example rule and its equivalent SQL code:

\vspace{1em}
\begin{lstlisting}[language=tondir]
R1(a, s) :- R(a, b, c),(s=sum(b)).
\end{lstlisting}
\begin{center}$\transformsto$\end{center}
\begin{lstlisting}[language=sql, style=sql_style]
WITH R1(a, s) AS { SELECT a, SUM(b) AS s FROM R }
SELECT * FROM R1
\end{lstlisting}
\vspace{1em}

Using this translation approach, a program will be translated to a chain of \codekw{WITH} clauses followed by a final \codekw{SELECT} statement that returns the results of the last translated rule.

% \smartpara{Final Renaming}
% To generate SQL, \system{} needs to consider possible collisions on relation/variable names. We present our collision resolution mechanism through an example. Suppose that we want to translate the following \xir{} to SQL:

% % \vspace{1em}
% \begin{lstlisting}[language=tondir]
% // Original schema of R is R(x,y) 
% R1(s) :- R(a, b), R(a, c), (c<10), (s=a*c).
% \end{lstlisting}
% % \vspace{1em}

% In this example, we do a self-join on relation \code{R} while applying a filter on the right-hand side relation. Then, we have an aggregation that will be solely projected in the results. Now, we show the SQL code generated by \system{}:

% % \vspace{1em}
% \begin{lstlisting}[language=sql, style=sql_style]
% WITH R1(s) AS {
%     SELECT SUM(R1.y*R2.y)
%     FROM R AS R1, and R AS R2
%     WHERE R2.y < 10
% }
% \end{lstlisting}
% % \vspace{1em}

% As it can be inferred from the SQL code, \system{} first set unique aliases for each involved relation. Then, using those aliases, it will rename all variables within the query using this pattern: \code{[relation\_alias].[original\_colname]}. To keep the generated SQL simple, \system{} does not rename any relation/variable when there is only one relation inside the rule. The column names needed for the variable renaming are fetched from the database schema (cf. Section~\ref{irdesign}). This phase also reverts possible column name mismatches that joins might deliberately introduce (cf. Section~\ref{pandastoir}).  

\smartpara{Sort and Limit}
A Common Table Expression (CTE) in SQL is very similar to the concept of views, but it is temporarily defined in the context of a larger query. CTEs are defined using \codekw{WITH} clause in SQL, and similar to the views, they cannot keep the results ordered after an \code{ORDER BY} command. Since in analytical queries, it is usual to sort and pick only a subset of the results  (e.g., almost all of the TPC-H queries), our code generation must be considerate about \codekw{ORDER BY} and \codekw{LIMIT} clauses in the program. To have a semantic-preserving translation, we alter the \xir{} program such that separately-defined \codekw{ORDER BY/LIMIT} pairs are done within a single CTE. Moreover, the individual \codekw{ORDERBY} clauses are only allowed in the final rule of a program.

\smartpara{Unique ID Generation} To keep the semantics of the Pandas indexes in our translations, we generate a unique ID column for the input relations. To generate such columns (translation of \code{UID()} expression) in \xir(), we use the following window function in SQL:
\vspace{1em}
\begin{lstlisting}[language=sql, style=sql_style]
SELECT
    row_number() OVER(ORDER BY col) AS ID
FROM rel
\end{lstlisting}
\vspace{1em}

In this SQL command, the order of \code{col} is used as a reference to generate sequential unique IDs. If this argument is not set, the database systems usually use the order of the first column in \code{rel}. However, because of the non-deterministic behavior of query engines in creating IDs when no ordering column is passed, we only use \code{UID} in the first appearance of a relation in the code and carry the generated column down to the end of the pipeline. It guarantees our immediate access to the ID column of a relation when it is required. We also do not allow translation rules to remove such a column. In the optimization phase (cf. Section~\ref{efficiency}), we prune the unused IDs and make the code as simple as possible. The generated IDs can be reset in specific situations discussed in Section~\ref{efficiency}.

\begin{figure*}[!t]
    \centering
    \begin{tabular}{cc}
        \begin{subfigure}[t]{.4\linewidth}
        \begin{lstlisting}
a=x.merge(y, on='id').
  drop('id', axis=1).
  to_numpy()
b=np.einsum('ij,ik->jk',a, a)
        \end{lstlisting} 
        
        \caption{Original Python code.}
        \label{fig:e2eex:a}
        \end{subfigure}
        &
        \begin{subfigure}[t]{.55\linewidth}
                \begin{lstlisting}
v1=x.merge(y, on='id')
v2=v1.drop('id', axis=1)
v3=v2.to_numpy()
v4=np.einsum('ij,ik->jk',v3,v3)
        \end{lstlisting} 
        
        \caption{Python code after ANF.}
        \label{fig:e2eex:b}
        \end{subfigure}
        \\
        \begin{subfigure}[t]{.4\linewidth}
        \begin{lstlisting}[language=TondIR]
v1(ID, c0, c1) :- 
  x(ID, c0), 
  y(ID, c1).
v4_1(ID, c0, c1, c2, c3) 
  group(ID) :- 
  v1(ID, c0, c1), 
  v1(ID, c2, c3),
  (c0=sum(c0*c2)), (c1=sum(c0*c3)),
  (c2=sum(c1*c2)), (c3=sum(c1*cf3)).
v4_2(c0) :- (c0=[0, 1]).
v4_3(ID, c0, c1) :- 
 v4_1(ID, c0, c1, c2, c3), v4_2(c4),
 (c0=(if(c4=0, c0, c2))),
 (c1=(if(c4=1, c1, c3))). 
        \end{lstlisting}
        
        \caption{Translated \xir{} code.}
        \label{fig:e2eex:c}
        \end{subfigure}
        &
        \begin{subfigure}[t]{.55\linewidth}
        \begin{lstlisting}[language=sql, style=sql_style]
WITH v1(ID, c0, c1) AS {
    SELECT r1.ID, r1.c0, r2.c1
    FROM x AS r1, y AS r2 WHERE r1.ID = r2.ID },
 v4_1(ID, c0, c1, c2, c3) AS {
    SELECT r1.ID,
    SUM(r1.c0*r2.c0) AS c0, SUM(r1.c0*r2.c1) AS c1,
    SUM(r1.c1*r2.c0) AS c2, SUM(r1.c1*r2.c1) AS c3
    FROM v1 AS r1, v1 AS r2
    WHERE r1.ID = r2.ID GROUP BY r1.ID }, 
 v4_2(c0) AS { VALUES (0), (1) }, 
 v4_3(ID, c0, c1) AS { SELECT
  (CASE WHEN r2.c0=0 THEN r1.c0 ELSE r1.c2) AS c0,
  (CASE WHEN r2.c0=1 THEN r1.c1 ELSE r1.c3) AS c1,
  FROM v4_1 AS r1, v4_2 AS r2 } SELECT * FROM v6
        \end{lstlisting} 
        
         \caption{Generated SQL code.}
         \label{fig:e2eex:d}
        \end{subfigure}\\
    \end{tabular}
    \caption{Example of an end-to-end translation from Python to SQL.}
    \label{fig:end2end_example}
\end{figure*}

\smartpara{Backend Adaptation}
Although most relational database engines have adopted the standard SQL as their interface language, there are minor details, mostly in the interface of their external functions (e.g. string operations), that the code generator should be aware of to generate a target-compliant SQL code. \xir{} makes it easy to support different dialects of SQL required by DBMSes, without the need to change any other components of \system{}; we generate SQL code compatible with two modern database systems (cf. Section~\ref{expsetup}).

\subsection{Putting it All Together}

To conclude Section~\ref{expressiveness}, we present an end-to-end example, demonstrating the compilation from a hybrid (Pandas/NumPy) program to SQL (Figure~\ref{fig:end2end_example}). In this example, we first join relations \code{x} and \code{y} on their shared \code{id} column. Then, we drop the \code{id} and convert the result DataFrame to a NumPy array (\code{a}). Subsequently, we pass the created array as the operands to an \codekwpy{einsum} API call that computes a covariance matrix. 

As the first step, Python code will be transformed to its ANF version (Figure~\ref{fig:e2eex:b}). Then, the equivalent \xir{} code of the ANF program is generated (Figure~\ref{fig:e2eex:c}). As discussed in Section~\ref{irtosql}, \xir{} does not allow dropping the ID columns. Thus, the \codekwpy{drop} API call is ignored in the translation. Since we already have the ID column, the \codekwpy{to\_numpy} API call is also ignored. Finally, \system{} plans the \codekwpy{einsum} API call and generates \xir{} using the ES8 kernel (cf. Table~\ref{tbl:einsum-kernels}). In the last phase of the transformations, based on the rules and techniques discussed in Section~\ref{irtosql}, \system{} generates SQL code using the \xir{} representation of the program (Figure~\ref{fig:e2eex:d}).

\section{Efficiency}\label{efficiency}
In this section, we discuss the optimizations applicable to the \xir{} representation of a data science workload. We will evaluate the effects of these optimizations in Section~\ref{microbench}. 

\smartpara{Local Dead Code Elimination} Local optimizations are applicable to a single rule in \xir{}. Here is an example:

\vspace{1em}
\begin{lstlisting}[language=tondir]
R1(y) :- R(a, b), (x=a), (y=a*b).
\end{lstlisting}
\begin{center}$\transformsto$\end{center}
\begin{lstlisting}[language=tondir]
R1(y) :- R(a, b), (y=a*b).
\end{lstlisting}
\vspace{1em}

As it is shown, there is an assignment in the rule that its variable is not listed in the head of the rule. In such cases, we remove the redundant assignment. 

\smartpara{Global Dead Code Elimination} Global optimizations are applied to a \xir{} program beyond the scope of a single rule. Consider the following code in \xir{}:

\vspace{1em}
\begin{lstlisting}[language=tondir]
R1(a, b, c, d) :- R(a, b, c, d), (a<10), (c=d).
R2(a, s) group(a) :- R1(a, b, c, d), (s=sum(a)).
\end{lstlisting}
\begin{center}$\transformsto$\end{center}
\begin{lstlisting}[language=tondir]
R1(a, b) :- R(a, b, c, d), (a<10), (c=d).
R2(a, s) group(a) :- R1(a, b), (s=sum(b)).
\end{lstlisting}
\vspace{1em}

We see that the attributes in the positions of variables \code{c} and \code{d} are not used in the second rule but are listed on the head of the first rule. In this scenario, we eliminate these unnecessary attributes from the first rule.

\smartpara{Group-Aggregate Elimination} There are cases where we can eliminate group-by aggregations. Consider this example:

\vspace{1em}
\begin{lstlisting}[language=tondir]
R1(ID, s) group(ID) :- R(ID, a, b, c), (s=sum(b)).
\end{lstlisting}
\begin{center}$\transformsto$\end{center}
\begin{lstlisting}[language=tondir]
R1(ID, s) :- R(ID, a, b, c), (s=b).
\end{lstlisting}
\vspace{1em}

In this example, there is a group-by-summation on the primary key of the relation \code{R1}. In such cases, if we know that the grouping column(s) are unique, we can remove the grouping and all the summation columns. \system{} brings this knowledge from the annotations or database catalog during the Python to \xir{} translation (Section~\ref{irdesign}).

\smartpara{Self-Join Elimination} In certain situations, there are redundant self-joins. Consider the following example:

% \vspace{1em}
\begin{lstlisting}[language=tondir]
R1(z) :- R(a,b1,c1,d1), R(a,b2,c2,d2), (z=b1*c2).
\end{lstlisting}
\begin{center}$\transformsto$\end{center}
\begin{lstlisting}[language=tondir]
R1(z) :- R(a1,b1,c1,d1), (z=b1*c1).
\end{lstlisting}
% \vspace{1em}

In this example, we can eliminate the self-join since (1) it is on a unique column and (2) no filter is applied to either of the relations. In this case, all the information needed from the second relation can be obtained from the first one without any loss.

\smartpara{Rule Inlining} By applying this optimization, we aim to fuse a chain of rules until we reach specific rules that are not fusible. We call these specific rules \textit{flow breakers}.\footnote{Not be confused with pipeline breakers in query engines.} Thus, in the first step, we find the dependency among rules and make a dependency graph. Then, based on the structure of each rule and the information in the dependency graph, we find the flow-breaker rules. In Table~\ref{tbl:breakers}, we summarized different kinds of flow breakers. To preserve the program semantics, we reset the index column (using \codekw{UID}) in each flow-breaker rule. We also do the same when facing joins in the rules. An example of rule inlining is shown below:

\begin{table}[!t]
    \caption{List of the flow breakers used in rule inlining.}
    \centering
    % \scriptsize
    % \begin{scriptsize}
    \begin{tabular}{|l|l|} 
    \hline
    \textbf{Flow Breaker} &
    \textbf{Description} \\
    \hline \hline
        Aggregate & 
        \codekw{agg} term in the body \\ \hline
        
        Group By & 
        \codekw{group} clause in the head \\ \hline

        Distinct &
        \codekw{unique} term in the body \\ \hline

        Sort/Limit & 
        \codekw{sort} and \codekw{limit} clauses in the head  \\ \hline
        
        % Computed Column & 
        % \codekw{x=t} and \codekw{x} is used by the dependent rule \\ \hline

        % UID & 
        % \codekw{UID} term in the body \\ \hline

        Outer Join & 
        outer-join terms (cf. Section~\ref{pandastoir}) in the body
        \\ \hline

        % Reused Rules & 
        % a rule with more than one dependants \\ \hline

        Sink Rule & 
        final rule in the program \\ \hline

    \end{tabular}
    \label{tbl:breakers}
\end{table}

\begin{lstlisting}[language=tondir]
R2(b, c, d) :- R1(a, b, c, d), (a > 1000).
R3(b, d) :- R2(b, c, d), (c != "A").
R5(e, g) :- R4(e, f, g), (f > 100).
R6(b, g) :- R3(b, x), R5(x, g)
R7(b, m) group(b) :- R6(b, g), (m = max(g)).
\end{lstlisting}
\begin{center}$\transformsto$\end{center}
\begin{lstlisting}[language=tondir]
R7(b, m) group(b) :- 
  R1(a, b, c, x), (a > 1000), (c != "A"), 
  R4(x, f, g), (f > 100), (m = max(g)).
\end{lstlisting}

Back to our end-to-end translation example, now we see the opportunity for applying group-aggregate elimination, self-join elimination, and rule inlining on the \xir{} representation of the program in Figure~\ref{end2end}.
\section{Evaluation}\label{evaluation}

\begin{figure*}[t]
\centering
\includegraphics[width=0.98\textwidth]{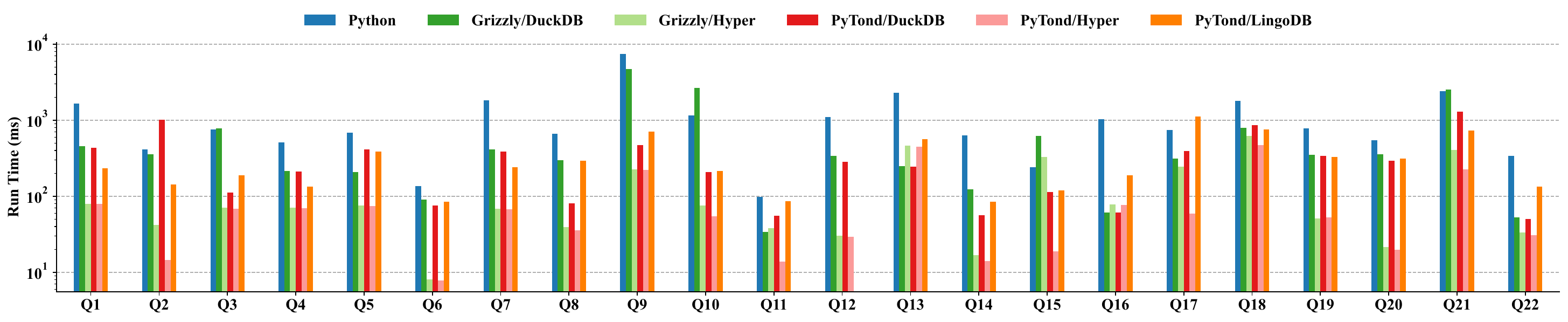}
\caption{Performance results for TPC-H workloads on a single thread.}
\label{fig:tpch_1}
\end{figure*}

\begin{figure*}[t]
\centering
\pagebreak{}
\includegraphics[width=0.98\textwidth]{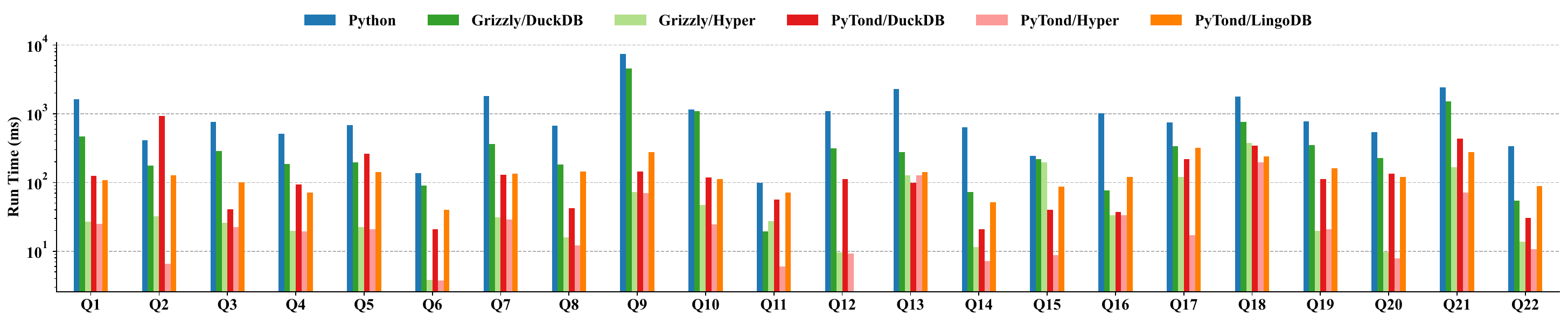}
\caption{Performance results for TPC-H workloads on 4 threads.}
\label{fig:tpch_4}
\end{figure*}

In this section, we evaluate our approach using different workloads and assess its performance compared to competitors. After presenting the experimental setup, we discuss the results of our end-to-end benchmarks. Then, through micro-benchmarks, we analyze the scalability of \system{} and discuss the impact of different optimizations.

\subsection{Experiments Setup}\label{expsetup}
To conduct the experiments, we used a single machine equipped with 16GB of DDR4 RAM and an Intel Core i5 1.6GHz with four cores and 256KB, 1MB, and 6MB of L1, L2, and L3 cache, respectively. Hyper-threading is disabled. We run experiments on Ubuntu 20.04 and use Python 3.10.13, NumPy 1.25.2, and Pandas 2.2.0. To compare the effects of different database engines as our backends, we use two modern database engines with different query execution paradigms. The first one is DuckDB~\cite{raasveldt2019duckdb}, a column-based vectorized in-memory database engine embedded in Python. The second one is Hyper~\cite{neumann2011efficiently}, a column-based in-memory engine representative of compiled query engines in our experiments. As our third backend option, we use LingoDB~\cite{DBLP:journals/pvldb/JungmairKG22}, an approach for efficient SQL to LLVM compilation (cf. Section~\ref{related}). Since LingoDB is a research prototype, we show its results whenever the SQL workloads can be executed correctly on this engine. Specifically, we exclude the Grizzly/LingoDB alternative as it lacks the support for certain SQL window functions required for UID generation (cf. Section~\ref{pandastoir}). We use DuckDB Python API 0.9.1, Hyper Tableau API 0.0.17782, and LingoDB Python API 0.0.1. We did five warm-up rounds in all the experiments and reported the mean of the next five execution rounds.

\begin{figure*}[t]
\centering
\includegraphics[width=0.98\textwidth]{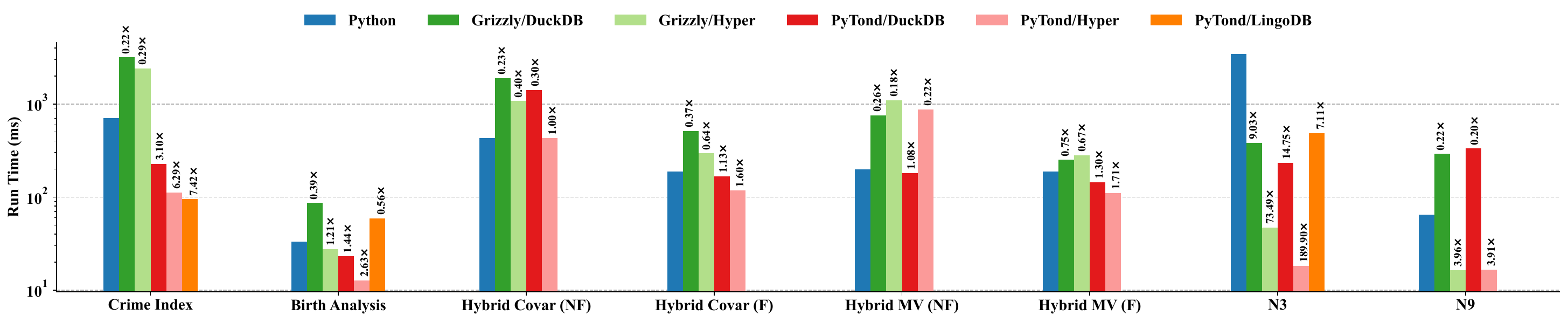}
\caption{Performance results for data science workloads on a single thread.}
\label{fig:others_1}
\end{figure*}

\begin{figure*}[t]
\centering
\includegraphics[width=0.98\textwidth]{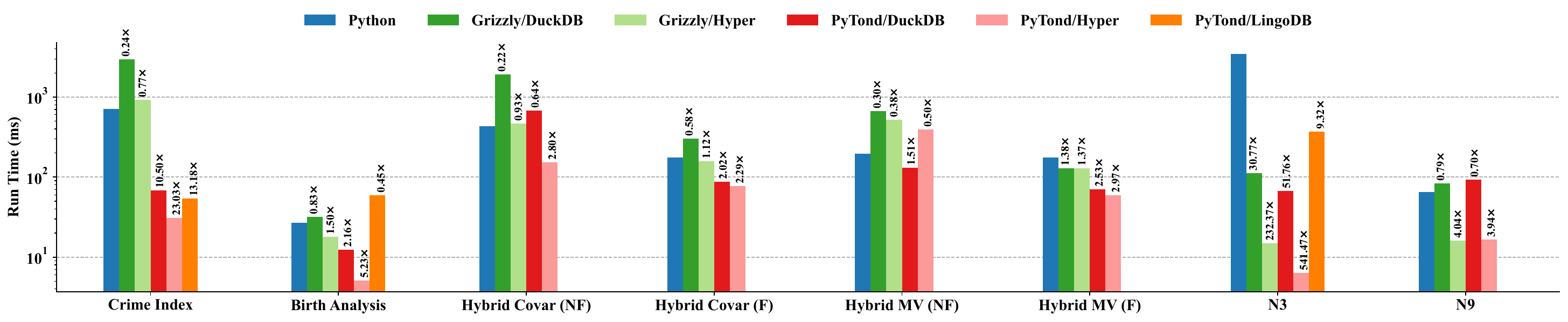}
\caption{Performance results for data science workloads on 4 threads.}
\label{fig:others_4}
\end{figure*}

\smartpara{Workloads}
We included a range of experiments to show the applicability of our approach to various data science workloads. We use the dense layout (cf. Section~\ref{expressiveness}) for linear algebra operations since it is (1) similar to the data layout to which the original Python code has access and (2) limitedly investigated by the current approaches~\cite{blacher2022machine, blacher2023efficient}. The characteristics of each approach are defined below.

\begin{itemize}
    \item \textit{TPC-H Queries:} TPC-H benchmark~\cite{tpch} is well-known in the analytical query processing. We use the Pandas version of TPC-H queries from~\cite{shahrokhi2023efficient} and run the queries on the TPC-H dataset with a scaling factor of 1. Although this benchmark does not represent the wide spectrum of data science workloads, it makes us confident about the system's ability to capture and process relational algebra, which is the backbone for most of the data analytics tasks.
    
    \item \textit{Crime Index:} We include the Crime Index data science notebook~\cite{palkar2018weld} (with the scaling factor of 100)  as a hybrid workload containing a sequence of Pandas, NumPy, and again Pandas operations. It filters an input DataFrame, converts it to an array, computes an \codekwpy{einsum}, converts the result of \codekwpy{einsum} back to a DataFrame, and returns the results after further filtering and projections.
    
    \item \textit{Birth Analysis:} This is another hybrid data science notebook~\cite{palkar2018weld} that contains NumPy \textit{Fancy Indexing} and Pandas \codekwpy{pivot\_table} APIs. We used the available statistical dataset for this experiment.
    
    \item \textit{Kaggle Data Science Notebooks:} We also bring two top-voted Kaggle data science workloads (N3 and N9) into our evaluation. The details of these benchmarks are discussed in PyFroid~\cite{emani2024pyfroid}. We used the computational parts of the notebooks (non-visualization parts) for the experiments. 
 
    \item \textit{Hybrid Matrix Calculation Experiments:} Besides the real-world data science workloads introduced before, this category adds two synthetic hybrid workloads. In both workloads, we first use Pandas to join two large tables. Then, we convert the result to a NumPy array. Finally, we execute an \codekwpy{einsum} over the results. In the first experiment, the \codekwpy{einsum} operation is a matrix-vector multiplication, while in the second one, it is a covariance matrix computation. We also include a slightly modified version of each experiment (\textit{Filtered} version) where a join-dependent filter is applied to the results of the initial join, before doing the final \codekwpy{einsum} calculation.
\end{itemize}

\smartpara{Alternatives and Competitors}\label{altandcompt}
To evaluate the performance of our approach, we included two competitors in our experiments. The first one is Python. We compare our in-database approach with the normal execution of workloads in Python, which is backed by Pandas and NumPy implementations. The second competitor is Grizzly~\cite{hagedorn2021putting}. However, as it will be discussed in Section~\ref{related}, its available version cannot provide the necessary support for our Pandas workloads. We assumed that Grizzly could cover our experimental workloads and consider the challenges in Python translation and SQL code generation (Section~\ref{expressiveness}). By putting this assumption, its generated SQL code would be similar to \system{}'s generated code before applying the optimizations. We use this simulated approach as our second competitor. There are barriers to including other competitors in our experiments. Firstly, except Grizzly, which we compare against, other very relevant approaches such as PyFroid~\cite{emani2024pyfroid} and Ponder~\cite{Ponder} lack accessible implementations for comparison. Secondly, certain competitors such as Daphne~\cite{DBLP:conf/cidr/DammeBB0BCDDEFG22} do not offer automated translations from Pandas/NumPy, necessitating different workload versions written in their specific APIs. These approaches are aligned with our goal of efficient data science but not with a focus on automated SQL code generation and its mentioned challenges. Alternatives such as Modin~\cite{petersohn2020towards} and Dask~\cite{Dask} are known for their high memory demand or distributed setup requirements~\cite{emani2024pyfroid}. However, \system{} aims to assist data scientists whether they are on commodity workstations or the cloud. Our preliminary analysis of Modin (on the same machine described in Section~\ref{expsetup}) shows poor performance (slower than Pandas) across TPC-H, Crime Index, and Birth Analysis benchmarks.

% \subsection{Experimental Results}\label{expresults}
% In this section, we present and discuss the results of our end-to-end benchmarks. Then, we show micro-benchmarks on some interesting aspects of the work.

\begin{figure}
\centering
\includegraphics[width=\columnwidth]{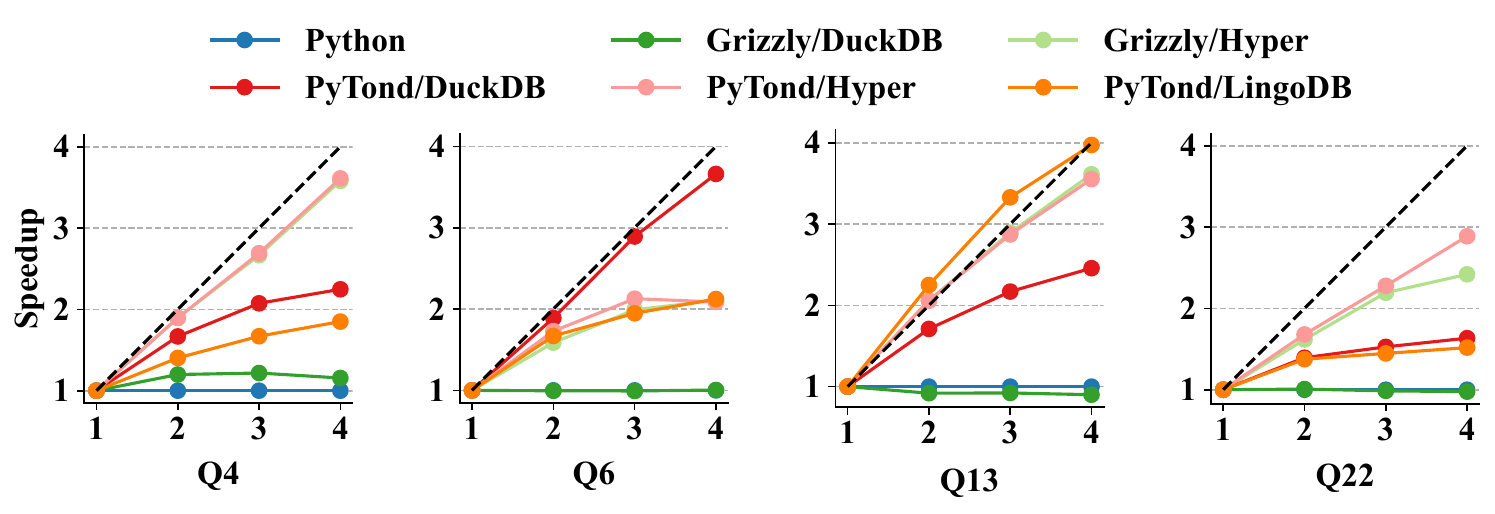}
\caption{Scalability analysis for representative TPC-H workloads on 1, 2, 3, and 4 threads.}
\label{fig:tcph_scaling}
\end{figure}

\begin{figure*}[t]
\includegraphics[width=0.97\textwidth]{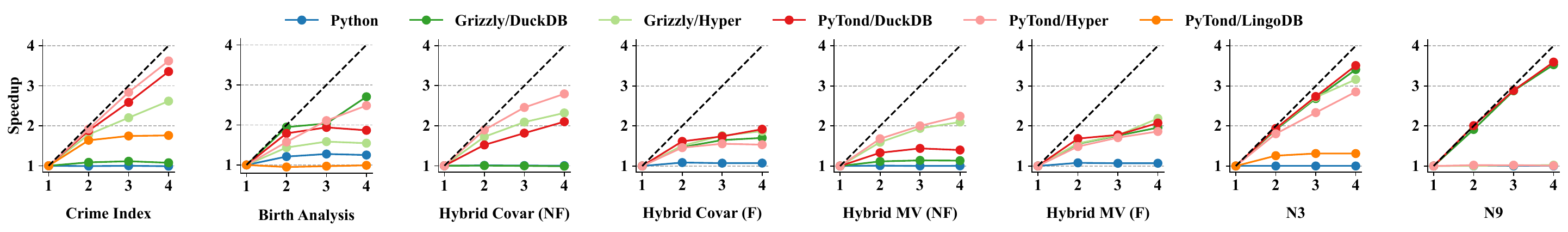}
\caption{Scalability analysis for hybrid workloads on 1, 2, 3, and 4 threads.}
\label{fig:others_scaling}
\end{figure*}

\begin{figure}
\centering
\includegraphics[width=\columnwidth]{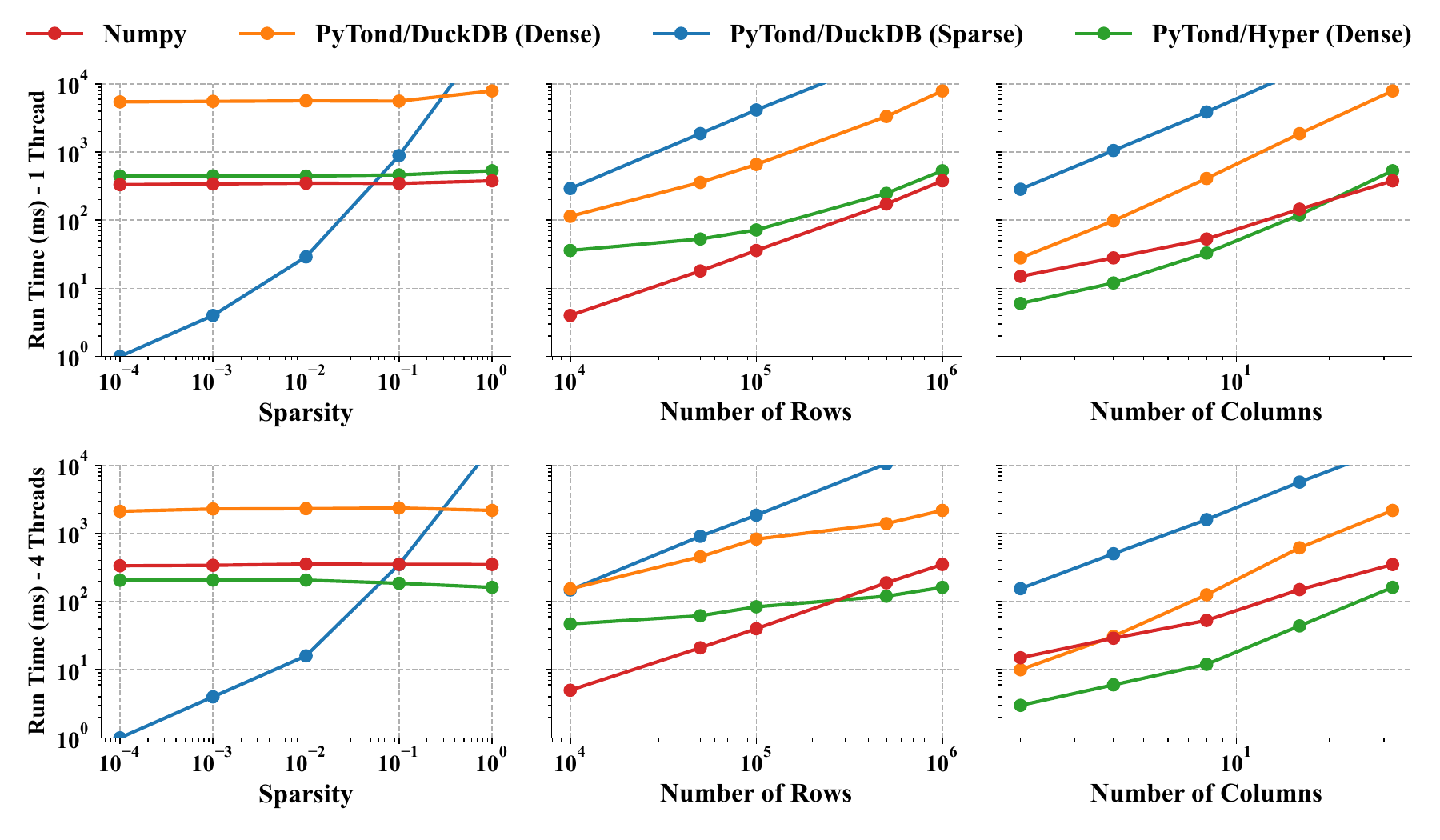}
\caption{Performance comparison of Covariance Matrix computation in NumPy vs \system{} with different backends and layouts. The \system{}/Hyper (Sparse) alternative is eliminated because of its incompetency. We cut the very large values for \system{}/DuckDB (Sparse) to improve the readability of the chart.  We used 1,000,000 rows, 32 columns, and a sparsity of 1 for the fixed dimensions. Both axes are on a logarithmic scale.    }
\label{fig:kernel}
\end{figure}

\subsection{End-to-End Benchmarks}\label{end2end}
In our end-to-end benchmarks, we measure the performance of all workloads when executed by our predefined alternatives. Since we assume that the data is already settled in the database (cf. Section~\ref{backgroundoverview}) and to do a fair comparison, we exclude the data loading time from all alternatives. The vertical axis in all Figures is in log-scale format.

We depicted the results of end-to-end experiments in the Figures~\ref{fig:tpch_1},~\ref{fig:tpch_4},~\ref{fig:others_1}, and~\ref{fig:others_4}. Figures~\ref{fig:tpch_1} and \ref{fig:tpch_4} show the results for all TPC-H queries on 1 and 4 threads, respectively. The initial observation drawn from these two charts is that PyTond is the first approach, offering complete coverage for the TPC-H benchmark. Figure~\ref{fig:tpch_1} also shows that for almost all queries, even the Grizzly-simulated approach (on different backends) performs better than the normal Python execution. It also shows that our approach can make the Grizzly-simulated approach much more efficient using its novel SQL rewriting. The effects of optimizations done in \system{} are more significant on the DuckDB backend, as we see a geometric average of 1.55$\times$ and 1.44$\times$ speedup on all queries after enabling them in DuckDB and Hyper. This implies that Hyper backend does more advanced query planning than DuckDB. Figure~\ref{fig:tpch_4} also verifies the same observations seen in Figure~\ref{fig:tpch_1}. Furthermore, it highlights that besides its inherent superiority over Python, our approach can benefit from parallelization, a feature not offered by Pandas. To summarize our TPC-H benchmarks, compared to Python and on a single thread, \system{} achieves a geometric average of 15$\times$ and 3.6$\times$ speedup on Hyper and DuckDB, respectively. It also achieves a geometric average speedup of 40$\times$ and 8$\times$ on the same engines on 4 threads. About the LingoDB backend, although its join processing could not process our generated SQL for Q12, in all other cases, its performance is similar to DuckDB.

The next two figures are related to our real-world and synthetic data science experiments. Successful execution of these experiments implies that our approach can capture Pandas/NumPy workloads with a variety of (complex) APIs such as DataFrame manipulations, \codekwpy{pivot\_table}, and \codekwpy{einsum}. Figure~\ref{fig:others_1} shows that \system{} is still competent or more efficient than Python and Grizzly-simulated approaches. These experiments reveal that even a systematic approach for Python to SQL translation (Grizzly-simulated) cannot offer high performance before applying the necessary optimizations. This can be seen in the relative performance of \system{} and Grizzly-simulated in all experiments, especially N3 and Crime Index. Further investigation on the effect of individual optimizations on the performance of \system{} is covered in Section~\ref{microbench}. Similar to Figure~\ref{fig:tpch_4}, Figure~\ref{fig:others_4} also verifies the same mentioned facts about the single-threaded setting of hybrid experiments while still showing the superiority of \system{} on 4 threads. An interesting result emerges from the N3 experiment, where \system{} achieves two orders of magnitude higher performance than Python. In this experiment, the Pandas data science code conducts a pipeline of relational algebra operations on 700MB of airline data. By analyzing the entire code, \system{} generates a compact and efficient SQL that executes the entire pipeline with minimal data copying and movement, leading to significant efficiency gains.

To investigate the performance of our approach in the context of linear algebra, we did a benchmark on Covariance Matrix computation, which is a fundamental algebraic operation used in machine learning. The results of these experiments are shown in Figure~\ref{fig:kernel}. In this benchmark, we take three primary properties of the input matrix including the number of rows, number of columns, and sparsity into account. In each chart, we fix two of these properties and vary the other one. The sparsity (left-most) charts clearly demonstrate that our sparse approach overtakes NumPy and other alternatives when the data is sparse. This is a significant achievement since the majority of machine learning models are sparse matrices. The other charts also imply that \system{} on Hyper is faster or competitive to NumPy with different input shapes.

\subsection{Micro-Benchmarks}\label{microbench}
This section shows the results of two different sets of micro-benchmarks. First, we assess the scalability of our approach on the number of threads. Then, we show the individual effects of each optimization on the \system{}'s overall performance. 

\smartpara{Scalability Analysis}
Figures~\ref{fig:tcph_scaling} and~\ref{fig:others_scaling} illustrate the scalability of Python, Grizzly-simulated approach, and \system{} on DuckDB, Hyper, and LingoDB backends. The scalability of each approach is shown in terms of speedup over its single-threaded performance. For brevity, we included a representative set of TPC-H queries~\cite{shahrokhi2023building} in our scalability experiments. These Figures show a smooth scale-up in \system{}'s performance specifically on DuckDB and Hyper backends. We can even find a near-linear scale-up in different cases including Q4, Q6, Q13, Crime Index, N3, and N9. In the Birth Analysis and N9 cases, the overall optimized run time is small, so the scalability results show saturated performance for some backends. In all of these experiments, there are two root causes behind Python's poor scalability. First, Pandas library does not support parallelization. Second, the more time-consuming part of hybrid workloads is attributed to Pandas code.

\begin{figure}[t]
\centering
\includegraphics[width=0.38\textwidth]{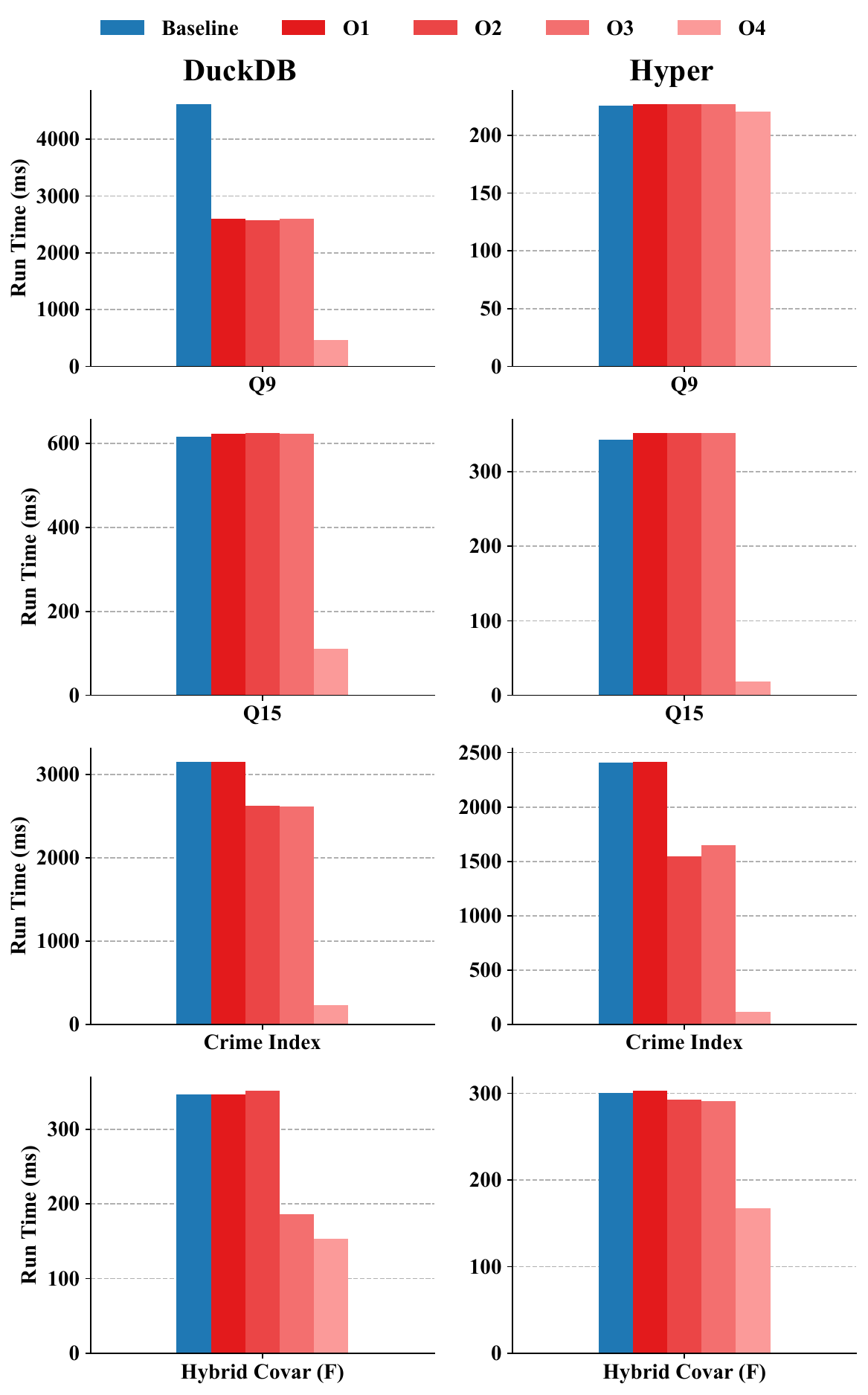}
\caption{Break-down of optimizations applied to experimental workloads. \textbf{O1}: Global/Local Dead Code Eliminations, \textbf{O2}: O1 + Group/Aggregate Elimination, \textbf{O3}: O2 + Self-join Elimination, \textbf{O4}: O3 + Rule Inlining}
\label{fig:opts}
\end{figure}

\smartpara{Effect of Optimizations} We analyzed the effects of optimizations for a representative subset of the workloads by starting from our baseline code (Grizzly-simulated) and applying each optimization on top of the others. The results are portrayed in Figure~\ref{fig:opts}. This figure implies that all of the optimizations introduced in Section~\ref{efficiency} are beneficial in their relevant scenarios. The \textit{O1} alternative shows the effects of \textit{Dead Code Eliminations} that improve the performance of Q9 in DuckDB. The \textit{O2} alternative adds the \textit{Group/Aggregate Elimination} and affects the Crime Index on both backends. This inefficiency is because of using a generalized \codekwpy{einsum} translation in the Crime Index that cannot be avoided (cf. Section~\ref{numpytoir}). The \textit{O3} alternative further improves the previous ones by applying \textit{Self-join Elimination}, which affects the Hybrid Covar on both backends. Finally, in \textit{O4}, we introduce \textit{Rule Inlining}, that its individual and combined effects result in huge performance improvements in most of the experiments.
\section{Related Work}\label{related}
In this section, we review state-of-the-art research on scalable data analytics in Python, discussing their characteristics and analyzing their strengths and weaknesses.

\smartpara{Compilation to Low-Level Code}
Solutions in this category \cite{shahrokhi2023building, lam2015numba, spiegelberg2021tuplex, palkar2018weld, klabe2022accelerating, DBLP:conf/cidr/DammeBB0BCDDEFG22} generate the low-level equivalent of Python code (e.g. in C++ or LLVM) that will be compiled for the target machine. SDQL.py~\cite{shahrokhi2023building} generates finely-tuned parallel C++ code for Pandas workloads by providing a domain-specific language (DSL)~\cite{DBLP:journals/pacmpl/ShaikhhaHSO22,DBLP:journals/pacmmod/SchleichSS23} that is \textit{embedded}~\cite{DBLP:journals/csur/Hudak96} in Python. Numba~\cite{lam2015numba} generates efficient low-level code for numerical operations written using the NumPy library. Numba users only need to add \code{@jit} decorator to their target functions to make them compiled. Tuplex~\cite{spiegelberg2021tuplex} and Klabe et. al~\cite{klabe2022accelerating} propose code generation for UDFs (User Defined Functions) in Python. Weld~\cite{palkar2018weld} generates LLVM code by lazy construction of an IR after each library API call. As opposed to Numba and \system{}, Weld users need to slightly modify the library APIs to make them adaptable to its compilation pipeline. Daphne~\cite{DBLP:conf/cidr/DammeBB0BCDDEFG22} provides an embedded DSL in Python for transforming integrated analytics to low-level code via MLIR. However, it lacks a compiler for translating Python data science code like Pandas and NumPy into its DSL, making user adaptation, benchmarking, and comparison difficult. \system{} provides interfaces for \textit{deeply} embedding~\cite{DBLP:conf/gpce/JovanovicSSNKO14} DSLs including Pandas, NumPy, and \xir{} itself in Python. This is achieved by adding a \code{@pytond} decorator to the functions that need to be compiled into SQL instead of low-level code.

Compilation of SQL queries to low-level code is a well-studied topic~\cite{Neumann11, dbtoaster, legobase_tods,dblablb, DBLP:journals/debu/ViglasBN14,DBLP:journals/jfp/ShaikhhaD018,DBLP:journals/pvldb/JungmairKG22}.
LingoDB~\cite{DBLP:journals/pvldb/JungmairKG22} introduces an MLIR-based framework that converts SQL into low-level efficient code. While they demonstrate a use-case of their approach with a PyTorch example, there is currently no available implementation of a Python interface for LingoDB for comparison purposes. Another related research~\cite{lingothesis} around LingoDB has only shown limited support of translation from Pandas (only three TPC-H queries), which contrasts with \system{}'s broad coverage.

The mentioned approaches have shown performance improvements by leveraging a variety of optimizations (e.g. parallelization and vectorization). Nevertheless, they introduce compilation overhead and potential portability issues. They also need users to declare input types, a requirement for low-level code generation, which can reduce the overall user-friendliness of the approach.

\smartpara{Compilation to SQL}
Approaches in this category~\cite{blacher2023efficient, blacher2022machine, hagedorn2021putting, fischer2022snakes, emani2024pyfroid, petersohn2020towards} aim to push the execution down into existing database engines by generating SQL code from source code. By taking this approach, the effort is mostly on a precise translation and then leveraging the proven optimization and execution power of databases. ByePy~\cite{fischer2022snakes} automatically generates SQL for generic imperative Python code with no support for its popular libraries such as Pandas and NumPy. Blacher et al.~\cite{blacher2022machine} proposed their translation methods based on training simple machine learning models (e.g., logistic regression). They experimentally show the merits of a database-friendly (dense) data layout versus the COO (sparse) one for numerical analysis, however, their approach is not automated. In another work, Blacher et al.~\cite{blacher2023efficient} presented a method to generate SQL for linear algebra workloads written in Einstein notation (\codekwpy{einsum} API in NumPy). While their approach can cover n-ary \codekwpy{einsum} and higher-order tensors, it can only operate on the data stored in a sparse format which, according to their experiments, is not always efficient. Grizzly~\cite{hagedorn2021putting} and PyFroid~\cite{emani2024pyfroid} generate SQL for Pandas. Our reconstruction efforts show that Grizzly APIs are not fully consistent with their Pandas equivalents and require the user to modify the source code. Moreover, Grizzly falls short in capturing different query operators, resulting in a limited coverage of standard TPC-H queries. It is also important to mention that its generated queries are not idiomatic SQL; they do not get the best performance from query engines. PyFroid~\cite{emani2024pyfroid} covers a broader set of APIs and incorporates optimizations on the generated code, the codebase of which is not accessible for experimentation.

\smartpara{Parallelization} Modin~\cite{petersohn2020towards} introduces a Dataframe algebra designed to encapsulate Pandas code and enable parallel execution. However, its reliance on Pandas as the backend limits achieving optimal performance on each core.

In this paper, by adopting the high-level strategy of the second category, we convert a wide range of Pandas (RA) and NumPy (LA) workloads to SQL. We use a decorator-based approach (like Numba~\cite{lam2015numba}) for source code translation to offer a good user experience. We rewrite (optimize) the generated SQL to fully exploit the capabilities of the query engine and let it do the parallelization, vectorization, and other advanced optimizations to efficiently execute the workload.
\section{Conclusion and Future Work}\label{conclusion}
Towards enhancing the efficiency of Python data science workloads, we presented \system{}, a framework for translating Python (Pandas/NumPy) to fine-tuned SQL that will be executed efficiently by database engines. We discussed the design of \xir{}, our intermediate representation, and elaborated on the required measures in translation, optimization, and code generation to make the most of DBMSes. Finally, we evaluated the performance of our approach in real-world scenarios, demonstrating its competency versus the alternatives. In the future, we aim to expand the work to capture more complex workloads with iterations~\cite{DBLP:journals/pacmmod/ShaikhhaSSN24, DBLP:conf/cgo/ShaikhhaSGO20}, recursions~\cite{DBLP:conf/sigmod/HirnG21, DBLP:conf/cidr/HirnG23}, and advanced numeric computations over higher-order tensors~\cite{DBLP:journals/pacmmod/SchleichSS23,DBLP:conf/cgo/ShaikhhaHH24}.

\section*{Acknowledgement}
The authors would like to thank Huawei for their support of the distributed data management and processing laboratory at the University of Edinburgh. This work was funded in part by a gift from RelationalAI.

\bibliographystyle{IEEEtran}
\bibliography{refs.bib}

\end{document}